\documentstyle[12pt,psfig]{article}
\def\reference{\parskip 0pt\par\noindent\hangindent 0.5 truecm}

\def\degr{\hbox{$^\circ$}}

\def\farcs{\hbox{$.\!\!^{\prime\prime}$}}
\begin{document}
%
% Title
% Capitalise the title normally - do not use ALL CAPS.
%
\title{Weak CSS Sources from the FIRST Survey}
%

% Authors
% Here comes the author(s) of the paper. Please add the appropriate author
% names for your paper and indicate within the $^...$ the number(s)
% which corresponds to the institute(s) of each author. In this example
% the second author has two institutional affiliations.
% Add or remove authors as required.
% **** IMPORTANT: Leave the closing curly bracket line as is. ******

\author{Andrzej Marecki$^{1}$,
 Jacek Niezgoda$^{1}$,
 Jacek  Wlodarczak$^{1}$,\\
 Magdalena Kunert$^{1}$,
 Ralph E. Spencer$^{2}$ and
 Andrzej J. Kus$^{1}$
} % IMPORTANT: leave this curly bracket as the first character of this line.

% Date - leave this blank.
\date{}
\maketitle

% Institutions
% Here fill in your institute name(s) and address(es)
% The number in $^...$ indicates the author number.  For example
{\center
$^1$ Toru\'n Centre for Astronomy, Nicholas Copernicus University, Toru\'n, Poland\\[3mm]
$^2$ Jodrell Bank Observatory, University of Manchester, UK\\[3mm]
}

% Abstract
% Simply place your abstract between the \begin{abstract} and
% \end{abstract} commands.
%
\begin{abstract}

We report early results of an observational campaign targeted on a sample of compact steep spectrum sources
selected from the FIRST survey which are significantly weaker than those investigated before. The selection
criteria and procedure are given in detail. We present here an assortment of MERLIN and VLBI observations and make
some general comments based on the morphologies of the sources presented.

\end{abstract}

{\bf Keywords:} surveys --- radio continuum: general --- quasars: general --- galaxies: active
% Place keywords here. Please write all keywords in lower case. PASA uses the
 %standard list of subject
% headings adopted by The Astrophysical Journal and available from URL:
%   http://www.journals.uchicago.edu/ApJ/keywords_text.html

% A formatting command to add space between the author list and the body
% of the paper when printed. This spacing may be changed as desired.
\bigskip

\section{Sample Selection}

To select weak compact steep spectrum (CSS) sources from the FIRST catalogue (White et al. 1997) we took the
following steps:

\begin {enumerate}

\item [a)]From the source list based on Green Bank (GB) surveys at 21 and
6\,cm (White \& Becker 1992) we selected those lying within the limits of the very first release of the FIRST
survey
--- it covered the area RA (J2000) $=6^h 35^m - 17^h30^m$, dec (J2000) $=28\degr - 42\degr$ at that time
--- having steep spectra ($\alpha>0.5$, $S\propto\nu^{-\alpha}$) and being stronger than 150\,mJy at 6\,cm. This
flux density limit was chosen in order to produce a sample of a manageable size. The above declination limits
indicate that the overlap between our sample and the B3--VLA survey-based sample (Fanti et al. 2001; Dallacasa et
al. 2002a, 2002b) is not large.

\item [b)] Thanks to a dramatic difference in the resolution, the majority
of sources appearing as single in the GB survey turn out to be double or multiple on FIRST maps. We rejected all
such cases, i.e.\ we selected only those sources that are single entities in the FIRST catalogue, i.e.\ more
compact than the FIRST beam (5\farcs4) and surrounded by an empty field. We adopted 1\,arcmin as a radius of that
field. Additionally we referred to NVSS (Condon et al. 1998) to check that, indeed, we picked up compact sources.
Such a procedure allows us to make sure that we deal with isolated objects and not parts of larger objects.

\item [c)]We again checked whether our targets fulfill the spectrum
steepness criterion: instead of GB survey flux densities at 21\,cm we used more accurate values from FIRST. We
rejected candidates with flat spectra ($\alpha \leq 0.5$).

\item [d)] We rejected the GPS sources. For this purpose we identified our
preliminary candidates with objects listed in the 365\,MHz Texas catalogue (Douglas et al.\ 1996). We passed only
those objects which have non-inverted spectra between 365 and 1400\,MHz. In other words the turnover frequencies
of our sources lie below 365\,MHz.

\end{enumerate}

Finally we selected 60 candidates for CSS sources.

\section{The Observations}

The initial survey was performed with MERLIN at 5\,GHz. Each of our targets was observed six times in 10 minute
scans spread evenly over a 12 hour track. Phase calibrators from the MERLIN Calibrator List were observed twice
per target scan for 1--2 minutes.

Six objects having sizes similar to classical CSS sources, yet being less luminous, were followed up with MERLIN
at 1.6\,GHz --- see Kunert et al. (2002).

Twelve targets have either been misdetected or unresolved. The latter group of nine objects have just been
followed up with the global VLBI. If it turns out that there are compact symmetric objects (CSOs) among them, this
will be a very interesting result. Normally CSOs are GPS sources but our sources are not GPS by definition so they
could be interpreted as `dying CSOs'. We comment on this group in Marecki, Spencer, \& Kunert (2003, hereafter
Paper II).

The whole set of MERLIN images is presented by A.\ Marecki et al. (in preparation). Twenty objects have been
selected for higher resolution observations. Each of them was observed with the VLBA at 18\,cm and the EVN at
6\,cm. Comprehensive results of those VLBI campaigns will be presented elsewhere. Here we draw some general
conclusions based on the whole observational material and illustrate them with a small subset of images.

\section{Some General Conclusions}

\begin{figure*}
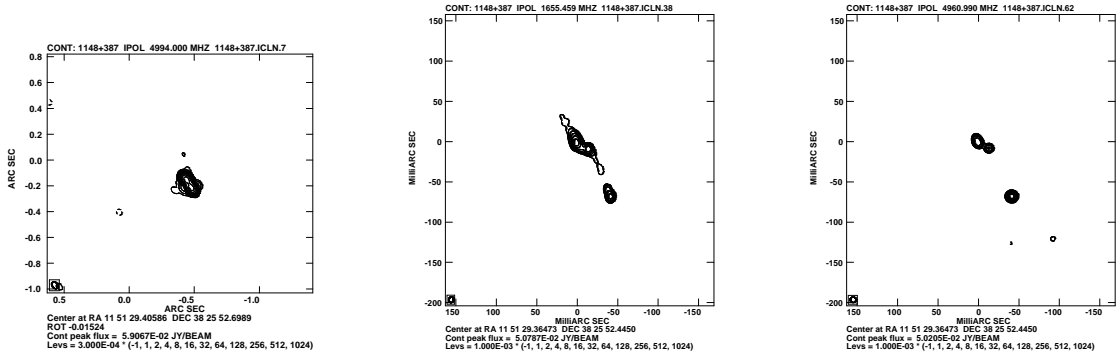

\vspace{-0.5cm} \hbox{\hspace{-0.8cm} \psfig{figure=1148+387.ICLN.ps,width=4.2cm}\hspace{1.0cm}
\psfig{figure=1148+387.ICLN18.ps,width=4.2cm}\hspace{1.0cm} \psfig{figure=1148+387.ICLN6.ps,width=4.2cm} }
\caption{\label{1148+387} MERLIN (6\,cm), VLBA (18\,cm), and EVN (6\,cm) maps of 1148+387.}
\end{figure*}

\begin{figure*}
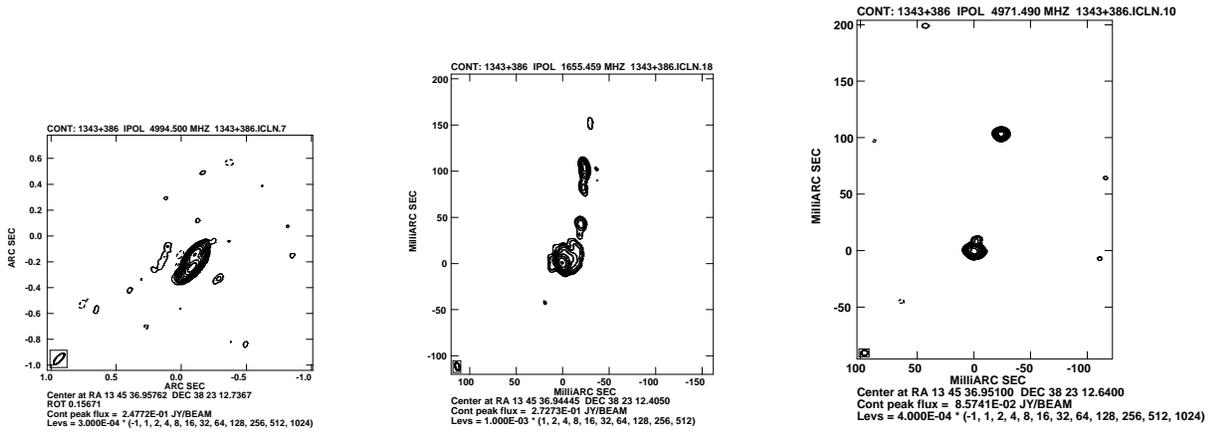

\vspace{-0.5cm} \hbox{\hspace{-0.8cm} \psfig{figure=1343+386.ICLN.ps,width=4.2cm}\hspace{1.0cm}
\psfig{figure=1343+386.ICLN18.ps,width=4.2cm}\hspace{1.0cm} \psfig{figure=1343+386.ICLN6.ps,width=4.2cm} }
\caption{\label{1343+386} MERLIN (6\,cm), VLBA (18\,cm), and EVN (6\,cm) maps of 1343+386.}
\end{figure*}

\begin{figure*}
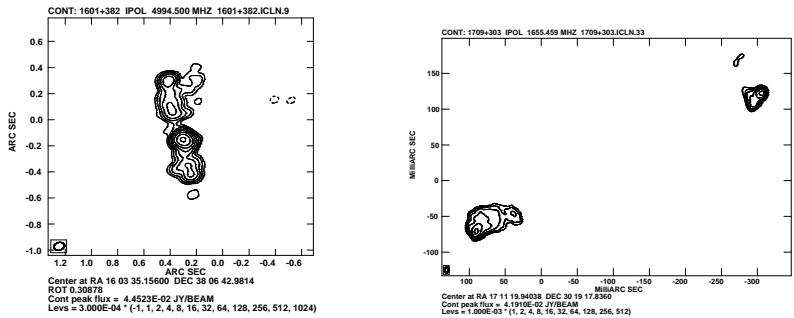

\vspace{-0.5cm} \hbox{\hspace{2.5cm} \psfig{figure=1601+382.ICLN.ps,width=4.2cm}\hspace{1.0cm}
\psfig{figure=1709+303.ICLN18.ps,width=5.2cm} } \caption{\label{FRIFRII} MERLIN (6\,cm) map of 1601+382 (left); VLBA
(18\,cm) map of 1709+303 (right).}
\end{figure*}

\begin{figure*}
\vspace{-1.0cm}
\hbox{
%\hspace{-0.8cm}
\psfig{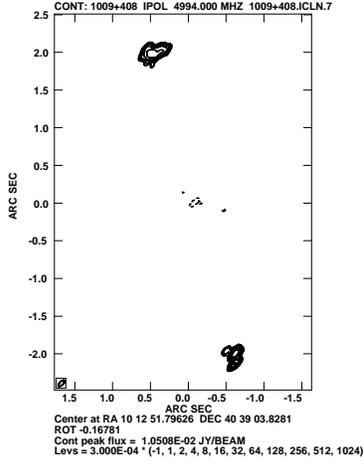}\hspace{2.0cm} \psfig{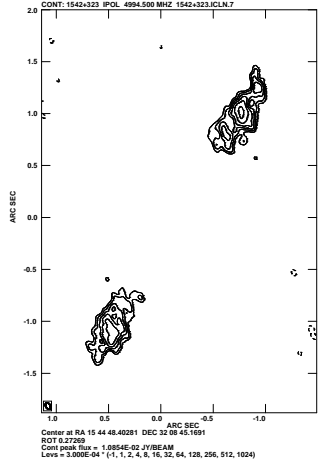} }
\vspace{0.5cm} \psfig{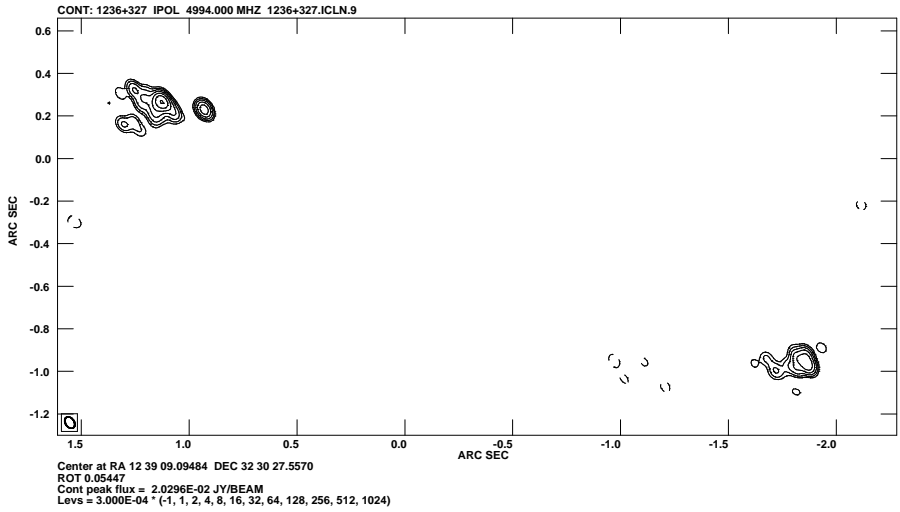} \caption{\label{dying_arcsec} MERLIN
(6\,cm) maps of `dying' arcsecond-scale CSS sources.}
\end{figure*}

\begin{figure*}
\vspace{-0.5cm}
\hbox{
%\hspace{-0.8cm}
\psfig{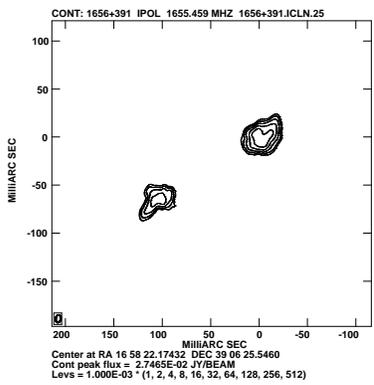}\hspace{1.0cm} \psfig{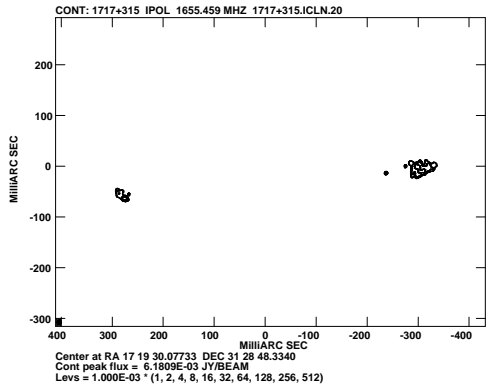} }
\caption{\label{dying_subarcsec} VLBA (18\,cm) maps of `dying' subarcsecond-scale CSS sources.}
\end{figure*}

\begin{enumerate}

\item Since our research was carried out at two frequencies and with
three different interferometric arrays it enabled us to see our targets in several frequency/resolution
combinations. As a result we were able to locate cores in many objects and it appears that one-sided structures
are {\em not} under-represented in our sample compared to other samples. In other words, it does not seem that CSS
sources in deep samples like ours are predominantly symmetric objects. A single network, single frequency map can
be very easily misinterpreted as an image of a symmetric object. The quasar 1148+387 is a good example of such a
deceiving object: it appears as a compact (0\farcs2 separation of the components) and very symmetric double on the
6\,cm MERLIN map but it is actually an asymmetric object according to our dual frequency VLBI surveys --- the
southern component has a flat spectrum so it must be a core (Figure \ref{1148+387}). 1343+386 (also a quasar),
which was  observed in parallel at 18\,cm with the VLBA by Dallacasa et al. (2002a), is a similar case (Figure
\ref{1343+386}). According to our measurements, the component Dallacasa et al. (2002a) denoted as N1 is the
strongest one and features a flat spectrum.

\item As in the case of other samples of CSS sources those which happen to
be medium-sized symmetric objects (MSOs) are FR\,II-like --- 1709+303 is a classical specimen --- and it is very
hard to find a mini-FR\,I; in our sample only 1601+382 resembles an FR\,I (Figure \ref{FRIFRII}). We further comment
on this fact in Paper II.

\item A classical FR\,II object (Fanaroff \& Riley 1974) is very clearly
edge brightened thanks to the hot spots located at the source's extremities, and CSS sources, regardless of
whether they are strong or weak, very often follow this pattern --- see e.g. 1709+303. However, we found a few
sources which, although in principle might be labelled FR\,II just because they are double-lobed but not FR\,Is, are
{\em not} featured by well-defined hot spots. These are 1009+408, 1236+327, 1542+323, 1656+391, and 1717+315
(Figures \ref{dying_arcsec} and \ref{dying_subarcsec}). Their lobes are diffuse and often `amoeba-shaped' ---
1542+323 (Figure \ref{dying_arcsec}) serves as the best example here. We think these are `dying' sources, i.e. AGN
where the central engine activity has stopped and the lobes are in the so-called `coasting phase'. This subclass
may be interpreted as a dead end of the evolution. We discuss this phenomenon in detail in Paper II.

\end{enumerate}

\section*{References}

% PASA uses the same conventions as ApJ for journal abbreviations.  Sample
% references are as follows.
% Please follow the same format for your references.

\reference Condon, J.J., Cotton, W.D., Greisen, E.W., Yin, Q.F., Perley, R.A., Taylor, G.B., \& Broderick, J.J.
1998, AJ, 115, 1693

\reference Dallacasa, D., Tinti, S., Fanti, C., Fanti, R., Gregorini, L., Stanghellini, C., \& Vigotti, M. 2002a,
A\&A, 389, 115

\reference Dallacasa, D., Fanti, C., Giacintucci S., Stanghellini, C., Fanti, R., Gregorini, L., \& Vigotti, M.
2002b, A\&A, 389, 126

\reference Douglas, J.N., Bash, F.N., Arakel Bozyan, F., \& Torrence, G.W., 1996, AJ, 111, 1945

\reference Fanaroff, B.L., \& Riley, J.M. 1974, MNRAS, 167, 31P

\reference Fanti, C., Pozzi, F., Dallacasa, D., Fanti, R., Gregorini, L., Stanghellini, C., \& Vigotti, M. 2001,
A\&A, 369, 380

\reference Kunert, M., Marecki, A., Spencer, R.E., Kus, A.J., \& Niezgoda, J. 2002, A\&A, 391, 47

\reference Marecki, A., Spencer, R.E., \& Kunert, M.\ 2003, PASA, 20, in press (Paper II)

\reference White, R.L., \& Becker, R.H. 1992, ApJS, 79, 331

\reference White, R.L., Becker, R.H., Helfand, D.J., \& Gregg, M.D. 1997, ApJ, 475, 479

\newpage

\newpage

\end{document}